\documentclass[aip]{revtex4-1}
\usepackage{graphicx}

\begin{document}


\title{Depth resolved grazing incidence time of flight neutron scattering from the solid-liquid interface} 

\author{M.~Wolff}

\email[]{max.wolff@physics.uu.se}




\author{J.~Herbel}

\author{F.~Adlmann}

\affiliation{Division for Materials Physics, Department of Physics and Astronomy, Uppsala University, Box 516, 751 20, Uppsala,
Sweden}

\author{A.~J.~C.~Dennison}

\affiliation{Division for Materials Physics, Department of Physics and Astronomy, Uppsala University, Box 516, 751 20, Uppsala,
Sweden}
\affiliation{Institut Laue Langevin, BP 156, 38042, Grenoble, France}

\author{G.~Liesche}

\affiliation{Hochschule Bremerhaven, Bremerhaven, Germany}
\affiliation{Institut Laue Langevin, BP 156, 38042, Grenoble, France}

\author{P.~Gutfreund}

\affiliation{Institut Laue Langevin, BP 156, 38042, Grenoble, France}

\author{S.~Rogers}

\affiliation{ISIS-STFC, Rutherford Appleton Laboratory, Chilton, Oxon OX11 0QX, United Kingdom}





\date{\today}

\begin{abstract}
We have applied small angle scattering in grazing incidence beam geometry on a time-of-flight neutron instrument.
Due to the broad wavelength distribution provided for a specific incident beam angle the penetration depth of the neutron beam is varied over a broad range in a single measurement.
The near surface structure of block co-polymer micelles close to silicon substrates with distinct surface energies are resolved.
It is observed that the very near-surface structure strongly depends on the surface coating whereas further away from the surface bulk like ordering is found.
\end{abstract}

\pacs{}

\maketitle 

\section{Introduction}

Neutrons have properties which enable them to be used as an unique probe in material research.
Their spin and low energy makes them sensitive to the magnetic induction in solids and suitable to investigate lattice vibrations and diffusive processes.
Most important in this context for the present study is that neutrons interact with the nuclei via the weak interaction.
As a consequence they are characterized by a weak absorption, resulting in a high penetration power, for many engineering materials like, e.g. silicon, aluminum or sapphire.
In addition the neutron is sensitive to different isotopes of the same element, which allows contrast variation experiments and highlighting of specific parts of a sample.
The two points just mentioned make neutrons an ideal probe for the study of buried liquid interfaces and surfaces by extracting the density profile in a reflectivity measurement.
In addition to the density profile along the surface normal, in-plane correlations are accessible via diffuse and small angle scattering.
As an example the ordering of micelles at the solid-liquid interface was studied using off-specular neutron scattering \cite{mcgillivray}.
Grazing incidence small angle neutron scattering (GISANS) expands on this technique by offering resolution for the scattering in all three dimensions \cite{wolff_1} thus providing a unique insight into the interfacial structure.\\
Following along this line recent developments in instrumentation have provided an optimized beam geometry and high neutron flux as required for GISANS studies \cite{chaboussant, mitamura}.
In this context experiments showed that the signal from the bulk of a sample might be separated from the surface scattering \cite{cousin} or that time-of-flight TOF-GISANS can be used to probe a large range of momentum transfers for a single incident beam angle \cite{kaune, buschbaum}.\\
One interesting class of materials to fully exploit the capabilities of this technique are block co-polymers since they have self-assembling properties on the molecular scale making them one promising way for the fabrication of nano-devices \cite{ludwigs, kim}.
In particular; close to a solid boundary one expects surface enrichment in case of an attractive interaction between the interface and one of the building blocks, or depletion in case of a repulsive interaction.
This may lead to lamellar phases with different orientations with respect to the interface \cite{huang}.\\
In aqueous solution block co-polymers may form micelles since the hydrophobicity of the building blocks may differ from each other resulting in the association with or exclusion of water molecules.
As the properties of different blocks may change with e.g. temperature or pressure, block co-polymers offer a model system for the study of gelation, percolation, crystallization or the glass transition \cite{chen}.
To become more specific for the system, consisting of polypropylene (PPO) and polyethylene (PEO) building blocks, presented here, increasing temperature results in a conformational change of the PPO towards lower polarity resulting in a loss of hydration of the polymer chains in water~\cite{GuoWangLiu1999}, whereas PEO blocks remain solvated across a broad range of temperatures.
At sufficiently high polymer concentrations this effect drives the macromolecules to aggregate into micelles with a hydrophobic core (PPO) surrounded by a more hydrophilic shell (PEO)~\cite{mortensen1996}.\\
In this article we describe a scattering experiment probing the near surface structure of a micellar polymer solution, which is in its cubic (fcc) phase, in contact to a solid boundary.
Our experiments go beyond previous studies, where depth sensitivity was shown for an air-sample interface with a thin layer deposited on the surface.
The present study is only possible since first $D_2 O$ has a smaller refractive index than silicon for neutrons and second silicon is nearly transparent for them.
This allows the investigation of the interface between two bulk materials in contact to each other.
We demonstrate that by combining a broad neutron wavelength spectrum with a grazing incidence beam geometry on a TOF-SANS instrument, the structure of polymer micelles can be resolved close to the solid surface as well as in the bulk in a single measurement.

\section{Experimental details}

\begin{figure}
\begin{center}
\includegraphics[width=13 cm]{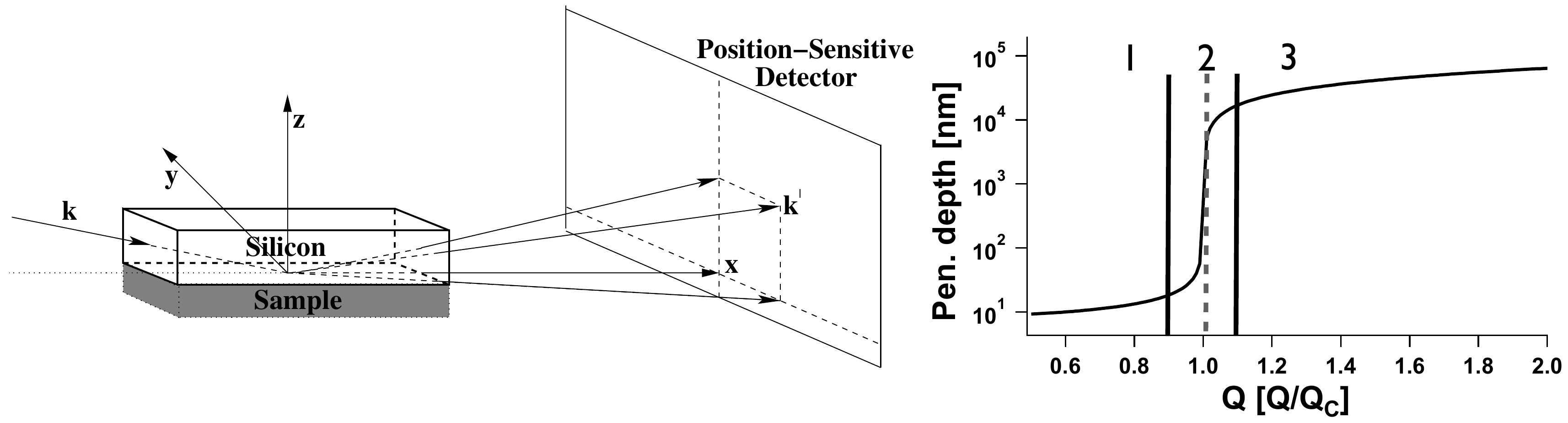}
\end{center}
\caption{\label{scattgeo}Schematics of the scattering geometry used for grazing incidence neutron scattering at the solid-liquid interface (left panel) \cite{wolff_1}. The right panel depicts the penetration depth of the neutron beam plotted over the transfer of momentum, normalized to the critical transfer of momentum for total external reflection.
The regions 1, 2 and 3 mark the Q intervals at which the GISANS data were taken.}
\end{figure}

Two functionalized single crystalline silicon (100) crystals (70*70*10 $mm^3$, polished, obtained from CrysTec, Germany) were used as solid substrates.
The roughness of the substrates was verified by x-ray reflectivity measurements and is below 0.5 nm.
One wafer was chemically cleaned in freshly prepared piranha solution (50/50 v/v $H_2SO_4$ (concentrated) and $H_2O_2$ (30 \% aqueous)) for 30 minutes and rinsed with Millipore water.
This procedure leads to an surface energy of approx. $\gamma =$ 72 $\frac{mJ}{m^2}$ at room temperature \cite{maccarini, gutfreund}.
The second wafer was cleaned in the same way and subsequently chemically covered with a self-assembled monolayer of octadecyltrichlorosilane OTS.
The resulting surface energy is about $\gamma =$ 19 $\frac{mJ}{m^2}$ at room temperature.
A detailed discussion of the roughness and surface energy measurements of the substrates used in the present study can be found in ref. \cite{gutfreund}.\\
The sample is a 18.5 \% (in weight, corresponding to 20 \% in $H_2 O$) solution of Pluronic F127 ((ethylene oxide)$_{99}$-(propylene oxide)$_{65}$-(ethylene oxide)$_{99}$) in $D_2 O$.
The bulk properties of this material have been reported in literature in great detail \cite{mortensen_1, mortensen_2}.
The polymer was purchased from Sigma-Aldrich and used without further purification.
The molecules were solved in $D_2 O$ (for better contrast) at a low temperature under constant stirring until a homogeneous solution was formed.
The sample cell was then filled with the sample in its liquid state at 5 $^\circ$C.
All measurements were performed at 25 $^\circ$C with the polymer solution in the crystalline phase.\\
Figure \ref{scattgeo} (left panel) depicts the scattering geometry for the neutron scattering experiments.
Neutrons penetrate a single crystalline block of silicon from the narrow side and are scattered at the solid-liquid interface.
A detailed discussion of the scattering geometry can be found in \cite{wolff_2}.
Note, measurements of solid-liquid interfaces are much easier using neutrons than x-rays, since the neutron has a high penetration power for many engineering materials, for example silicon, aluminum and steel.
This property is combined with a large scattering potential of light elements and in particular of deuterium, resulting in a finite angle of total external reflection at the silicon $D_2 O$ interface.\\
Table \ref{SLDQC} summarizes the scattering length density and electron density for silicon and the sample.
\begin{table}
\begin{center}
\begin{tabular}{c|c|c}
						&SLD Neutron				&SLD x-ray		\\ \hline \hline
Silicon					&2.08*10$^{-6}$~\AA$^{-2}$	&20.1*10$^{-6}$~\AA$^{-2}$     \\
Sample					&5.17*10$^{-6}$~\AA$^{-2}$	&9.35*10$^{-6}$~\AA$^{-2}$	\\
\end{tabular} 
\end{center}
\caption{\label{SLDQC} Neutron scattering length densities and electron densities for the interface studied in the present work.}
\end{table}
For x-rays no critical angle of total external reflection exists at the silicon-liquid interface.
On the other hand neutrons are totally reflected for a momentum transfer smaller than $Q_c=0.124$ nm$^{-1}$.
This implies that for an incident angle of 0.3$^{\circ}$ the critical wavelength for total external reflection is $\lambda_c=0.53$ nm.
Wavelengths longer than $\lambda_c$ are totally reflected with only the evanescent wave penetrating the sample, whereas shorter wavelength neutrons are refracted and transmitted into the liquid.\\
The TOF-GISANS experiments were performed on the instrument SANS-2D at ISIS (Rutherford-Appleton Laboratory, Didcot, England).
The choppers were set to allow a wavelength band of 0.175 - 1.56 nm which prevents frame overlap of preceding and subsequent pulses at the repetition rate of 10 Hz generated at TS2 at ISIS.
In order to probe different penetration depths the intensities were integrated for one incident beam angle over wavelength intervals of 0.175-0.5 nm, 0.5-0.6 nm, 0.6-1.56 nm, corresponding to wavelength resolutions of $\frac{\Delta \lambda}{\lambda}$ of 48 \%, 9 \% and 44\%, respectively.
In addition; the lattice parameter of the crystalline structure is evaluated for more narrow wavelength bands.
The incident beam was collimated over a distance of 6 m and had a horizontal and vertical divergence of 0.29$^\circ$ and 0.04$^\circ$ (base width of a triangular).
The source and sample slits were set to 20*4 mm$^2$ and 10*0.2 mm$^2$, respectively.
The footprint of the sample for the incident beam angle of 0.3$^\circ$ was about 0.25 mm to avoid over illumination and air scattering, which would result in additional background.
The detector with a pixel resolution of 5*5 mm$^2$ was set to a distance of 6 m.\\

\section{Results and discussion}

\subsection{Result}

\begin{figure}
\begin{center}
\includegraphics[width=13 cm]{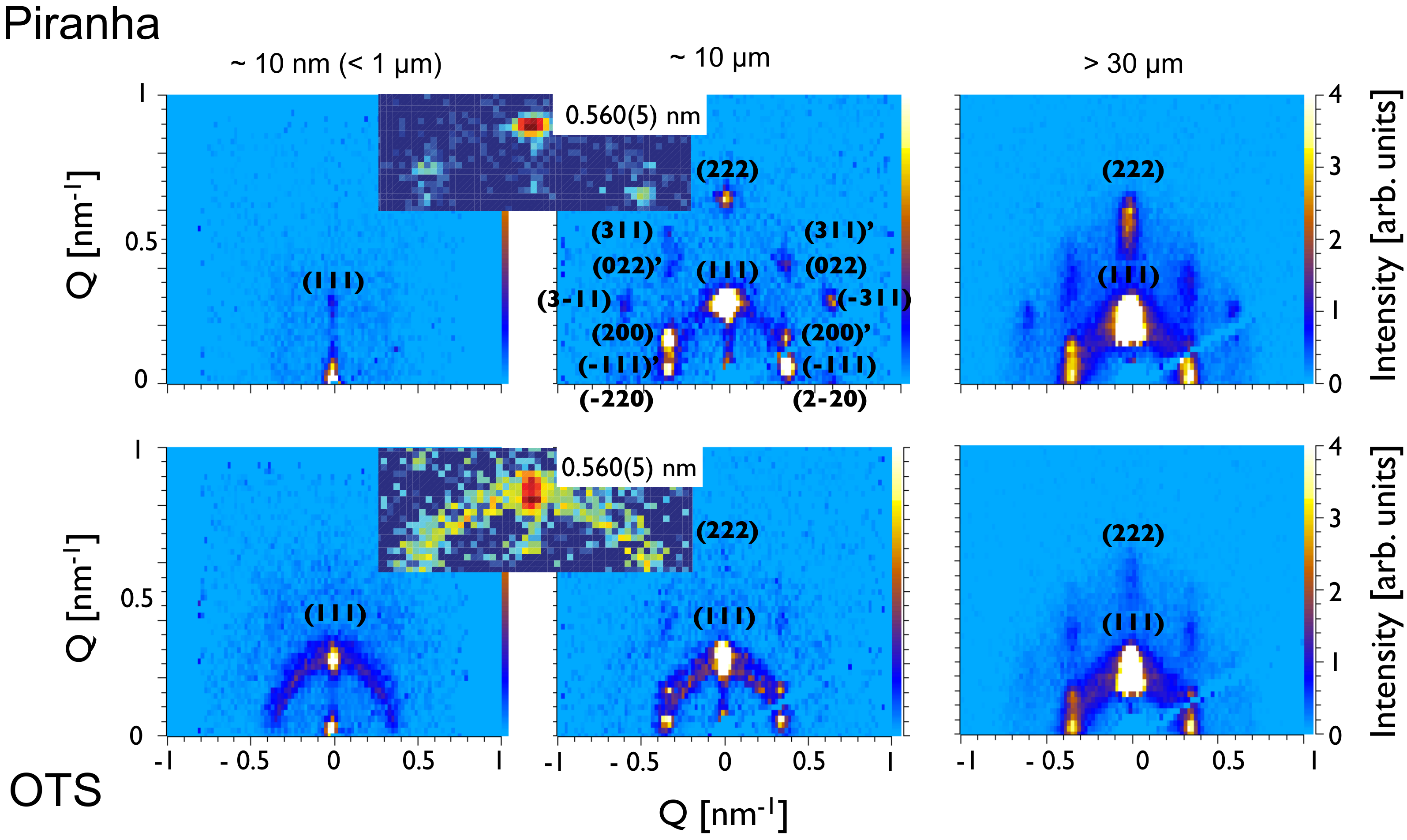}
\end{center}
\caption{\label{GISANS}GISANS data taken for an 18.5 \% (by weight) solution of the polymer F127 solved in $D_2 O$ in contact with a silicon surface treated with piranha solution (top panels) and OTS (lower panels).
The different penetration depths, noted at top and increasing from left to right, result from the different wavelengths in the incoming beam.
Crystalline ordering is clearly preferred in the vicinity of the piranha solution treated surface.}
\end{figure}

Figure \ref{GISANS} depicts the patterns of intensity measured on SANS-2D \cite{sans-2d} for an 18.5 \% (by weight) solution of the polymer F127 solved in $D_2 O$.
The three pictures for the silicon surface treated with piranha solution (top panels) and OTS (lower panels) were measured simultaneously for the incident beam angle of 0.3 degree with a fixed detector angle.
The direct as well as the reflected beam were masked by the beam stop in order to avoid detector saturation and reduce the background.
The left panels correspond to the integration of the detector images for wavelengths ranging from 0.6-1.56 nm resulting in a penetration depth of the neutron beam into the polymer solution of approximately 10 nm (region 1 in figure \ref{scattgeo}) or 1 $\mu$m taking into account the tail in divergence of the incident beam and marked by the grey dashed line in figure \ref{scattgeo}.
The intensity integrated for short wavelengths of 0.175-0.5 nm which have a large penetration depth of about 30 $\mu$m are shown in the right panels.
The central column summarizes the intensities for wavelengths integrated around the critical wavelength, 0.5-0.6 nm, resulting in an intermediate penetration depth of about 10 $\mu$m.\\
For the smallest penetration depth of about 10 nm a ring of scattered intensity is visible around the direct beam (Q=0) for the OTS substrate.
This ring is absent for the micelles in contact with the surface treated with piranha solution.
It should be pointed out that due to the divergence of the incident beam the angular spread of the incident beam results in probing a range of penetration depths; this varies from 10 nm to around 1 $\mu$m for those wavelength with a nominal penetration depth of 10 nm.
The presence of Bragg reflections at $Q_z \approx 0.4$ which corresponds to features larger than the nominal 10 nm penetration depth is explained by the effect of this divergence.
Clear qualitative differences between the scattering from the two samples can be seen in the central panels of fig. \ref{GISANS} where the penetration depth is nominally 10 $\mu$m.
10 well resolved Bragg peaks, together with 4 weaker peaks are observed for the sample close to the piranha treated surface, whereas for the OTS surface peaks of a lower intensity together with a Debye-Scherrer ring at Q = 0.4 nm$^{-1}$ are observed.
The insert between the left and middle panels depicts a zoom into the low Q-region and with a narrow wavelength band around 0.56 nm, which is slightly larger than the critical wavelength.
On the two panels the difference in structure between both interfaces is most striking.
For the piranha cleaned surface clear Bragg reflections are visible even with an asymmetry in the reflection found at positive and negative $Q_y$ values, respectively.
This implies a highly textured well ordered crystalline structure in the vicinity of the interface.
On the other hand for the OTS surface a ring of increased intensity is found corresponding to a powder or amorphous surface layer.
For the largest penetration depth (right panels) both detector images become more similar; for the piranha solution treated surface the diffuse scattering becomes increased, whereas for the OTS one the reflections are separated more clearly.

\subsection{Reciprocal space}

\begin{figure}
\begin{center}
\includegraphics[width=10 cm]{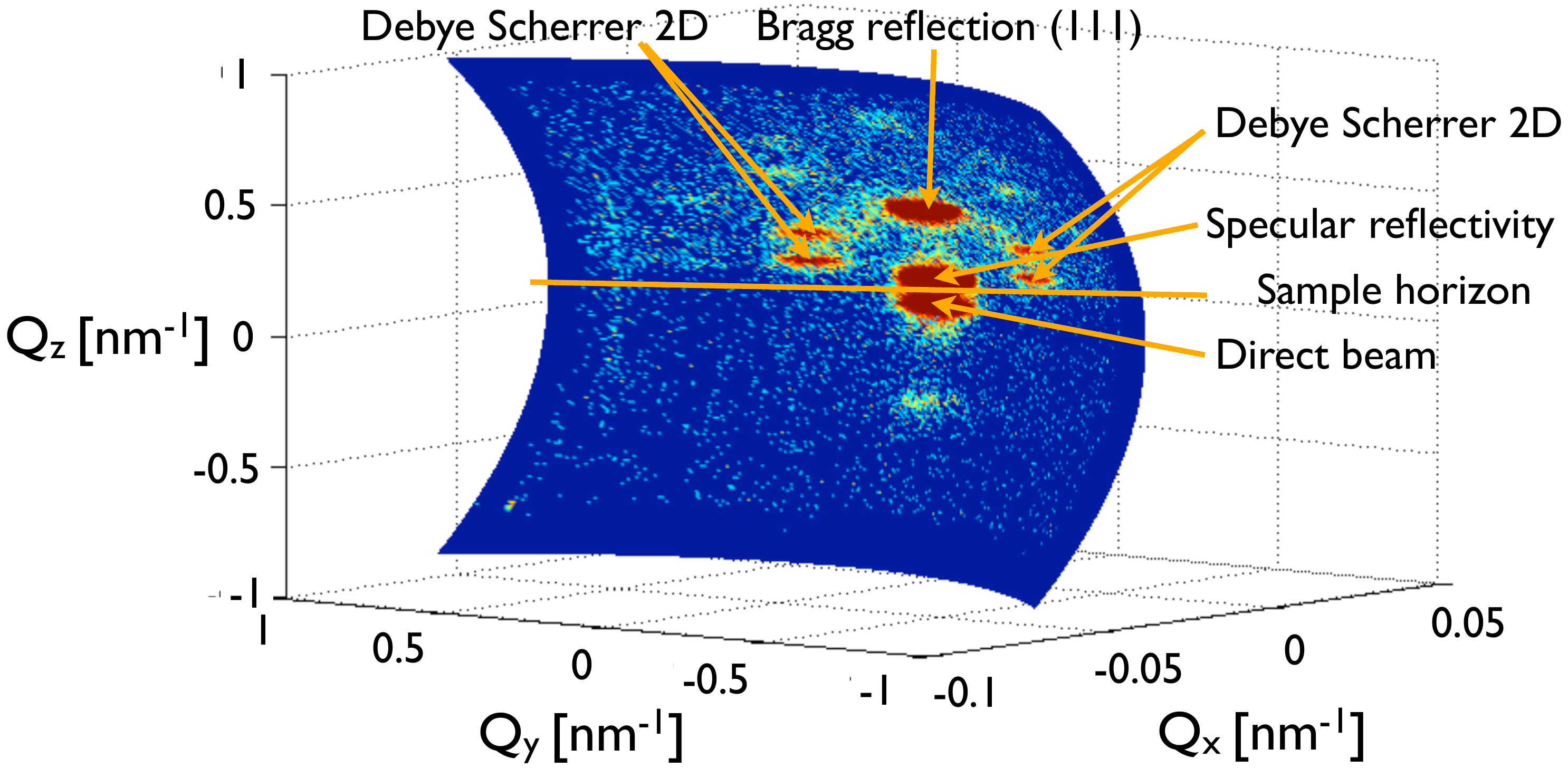}
\end{center}
\caption{\label{rec_space}Detector image for a fcc crystalline structure plotted as color map on the surface of the Ewald sphere.}
\end{figure}

In order to understand the crystalline structure formed at the different distances from the solid substrate and to relate those to the scattering patterns presented in Figure \ref{GISANS} it is important to have a closer look at the scattering geometry and to understand how the instrumental settings affect our observations.
Figure \ref{rec_space} depicts intensities for the same kind of cubic structure formed by F127 micelles solved in $D_2 O$ taken on the instrument ADAM at the Institute Laue-Langevin (Grenoble, France) \cite{wolff_3, devishvili}.
This instrument uses a monochromatic neutron beam of 4.41 \AA.
The incident beam was collimated in order to well resolve the Bragg reflections.
Figure \ref{rec_space} shows the scattering pattern at a single point near the critical reflection of the surface on the Ewald sphere.
The logarithm of the scattered intensity is plotted as a color map (high and low intensity in red and blue, respectively).
The coordinate system x, y and z is defined with respect to the sample surface as shown in Figure \ref{scattgeo}.
A more detailed discussion of the length-scales probed along the different directions can be found in \cite{wolff_4}.
It is clear that the detector plane represents a curved surface with Q=0 at the position of the direct beam.
The second and only other point where the in-plane coordinates $Q_x$ and $Q_y$ are zero is the point where the specular scattered intensity is detected.
All other points on the detector have none-zero $Q_x$ and/or $Q_y$ components.\\
Considering a densely packed cubic structure at the interface, the (111) lattice planes would be parallel to the solid-liquid boundary and the Bragg reflections resulting from them should have no in-plane component ($Q_x$ = $Q_y$ = 0).
This implies that the (111) reflection can only be detected if it has a certain rocking width along $Q_x$.
The values probed along the $Q_x$ direction are much smaller than those probed along $Q_y$ and $Q_z$ and for a crystal coherence on the order of $\mu$m a finite intensity is detected in the detector plane.
Moreover with decreasing wavelength the momentum transfer along the x direction decreases at the position of the Bragg refelection for small Q \cite{wolff_2}.
This implies that for a specific crystal coherence the intensity scattered into the detector plane should increase with decreasing wavelength.
All other four Bragg reflections are only visible if the crystalline structure is a two dimensional powder with respect to the sample interface \cite{wolff_4}, since the $Q_y$ values are much larger than the $Q_x$ values and the probability of a crystal orientation with the Bragg reflection on the Ewald sphere is unlikely.
Moreover, for a single crystal arrangement the reflections with and without prime visible in figure \ref{GISANS} can never be detected on the Ewald sphere at the same time, since they would be separated by an angle of 180$^\circ$ on the Debye-Scherrer ring, which is not possible for the symmetry of the crystal.\\

\subsection{In-plane correlations}

As mentioned in the previous section the (111) reflection can only be detected if it has a certain rocking width along $Q_x$.
This rocking width along $Q_x$ transfers directly into a in-plane correlation length by taking the Fourier transform from reciprocal to real space.
\begin{figure}
\begin{center}
\includegraphics[width=5 cm]{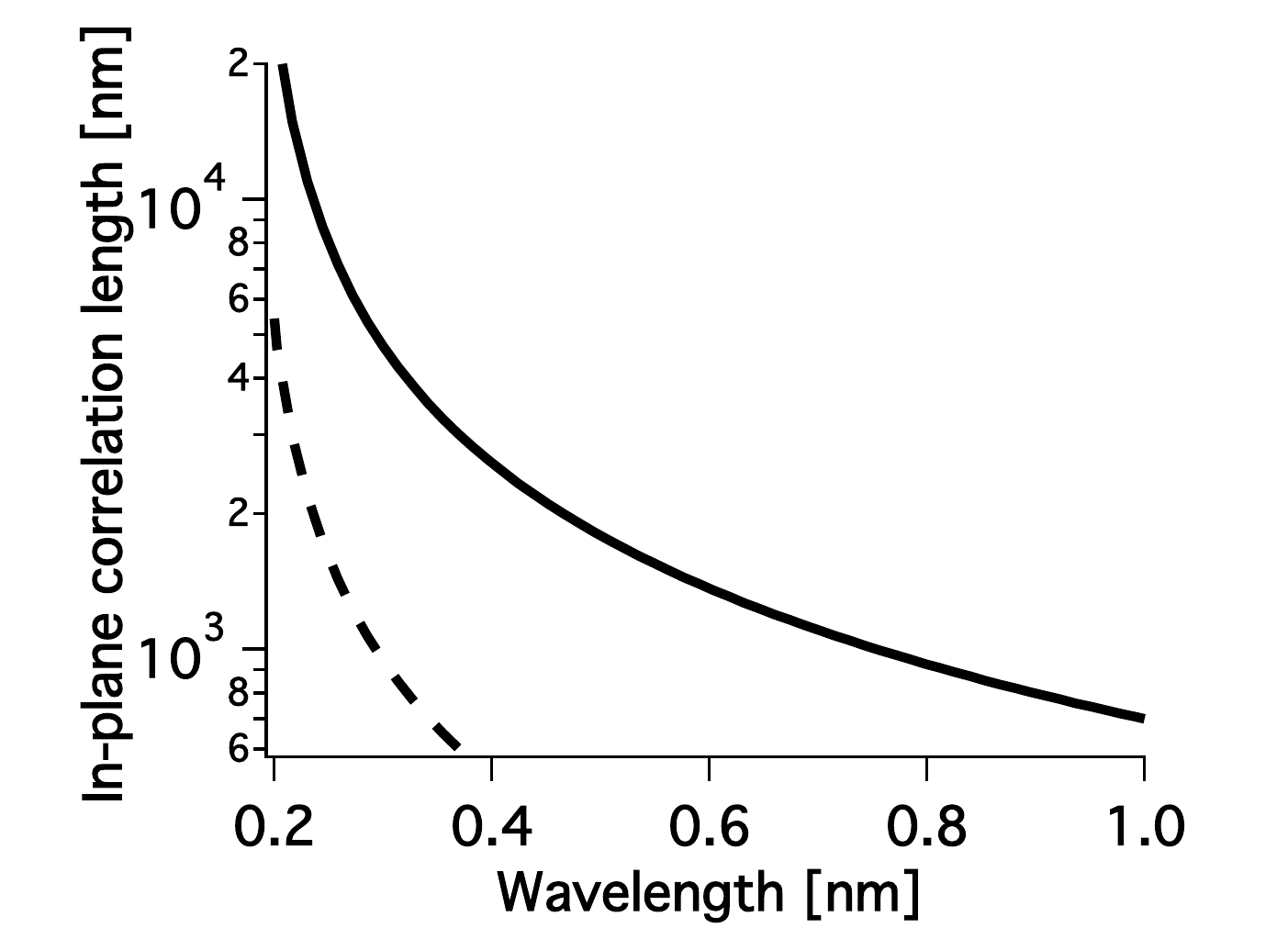}
\end{center}
\caption{\label{Correlation}In-plane correlation length probed for the different wavelength at the position of the (111) Bragg reflection.}
\end{figure}
Figure \ref{Correlation} shows how the correlation length varies as a function of wavelength at the (111) peak position.
The solid line represents a full width half maximum of the diffuse scattering corresponding to the $Q_x$ values probed in the detector plane.
The dashed line represents a line width which is five times larger.
A strong Bragg reflection is only expected in the area between the two lines.\\
Considering the scattering geometry the intensity distribution visible in Figure \ref{GISANS} is explained by the following model.
The crystal has a larger coherence and better epitaxy close to the surface cleaned with piranha solution.
For the smallest penetration depth (left panels) and the largest wavelength a relatively large $Q_x$ is probed.
Considering the large coherence of the crystal it becomes clear why almost no intensity is visible in the upper left panel.
On the other hand for the more powder like structure with small coherence close to the OTS surface the ring of intensity visible in the lower panel on the left hand side becomes understandable.
For decreasing wavelength and increasing penetration depth the component of $Q_x$ at the Bragg peak becomes smaller.
The good crystallinity next to the piranha solution cleaned surface is well reflected by the significantly increased intensity of the (111) reflection and the large number of visible reflections in this case.
For the OTS interface the intensity increases less and a lot of diffuse scattering is visible.
For the largest penetration depth (right panels) the two scattering patterns, top and bottom, become more similar resembling a more bulk like ordering in the sample.
In addition all reflections become more smeared out since the intensity is integrated over a large wavelength band.\\

\subsection{Lattice parameter}

\begin{figure}
\begin{center}
\includegraphics[width=7 cm]{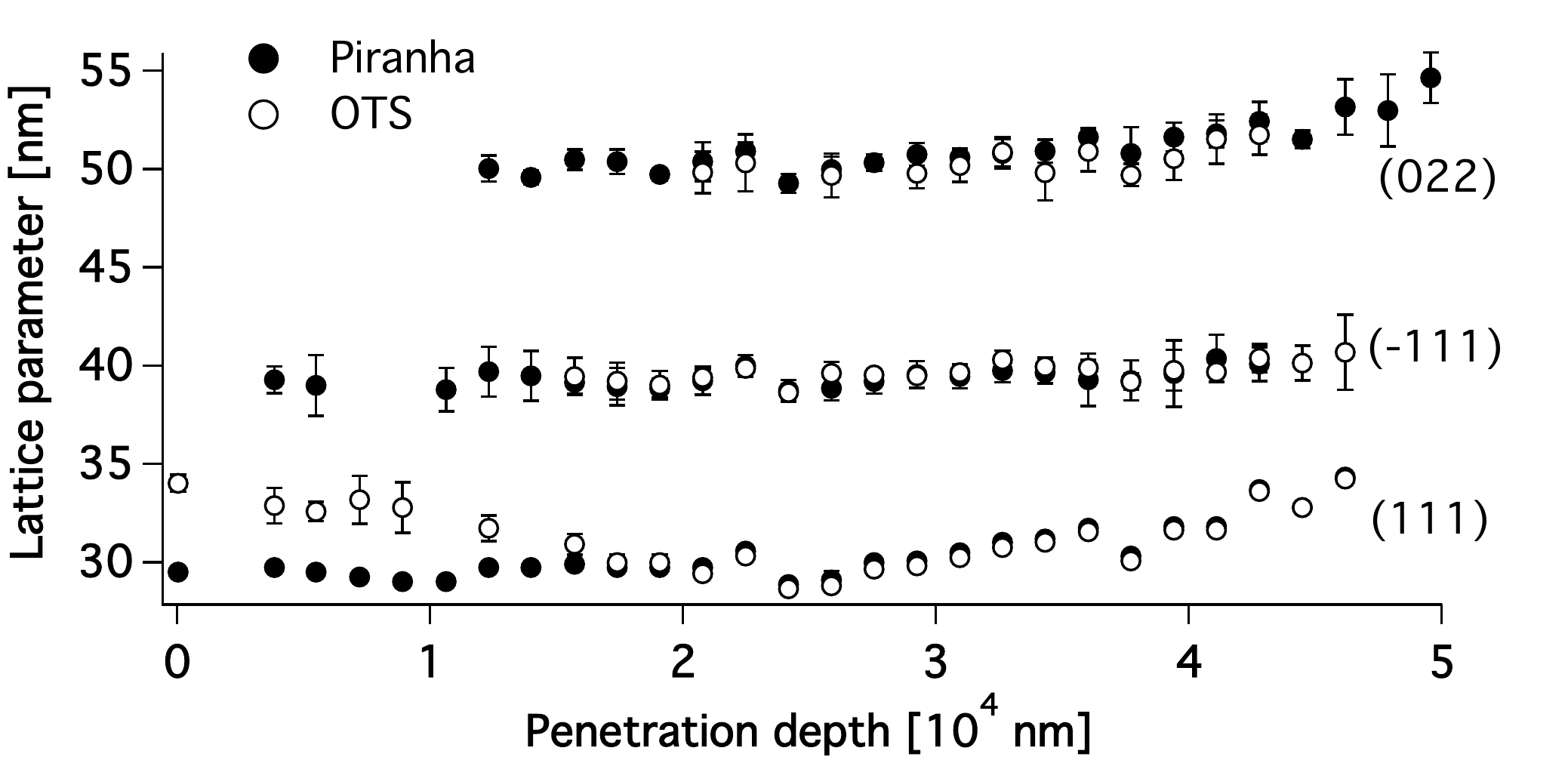}
\end{center}
\caption{\label{d-space}d-spacing extracted for different Bragg reflections and plotted versus penetration depth. For clarity, the data for the (-111) and (022) reflection are shifted by 10 and 20 nm, respectively.}
\end{figure}

One parameter which is relatively easy to extract from the data is the position of the different Bragg reflections.
As a result of this evaluation the lattice parameter of the fcc crystal structure is depicted in fig. \ref{d-space} for both samples.
For the data evaluation the intensity was integrated over a narrow wavelength band of typically 0.1 - 5 pm.
Subsequently, the intensity map is fitted by a two dimensional Gauss function and Q is extracted from the position of the peak maximum.
The error bars shown in the figure represent the mathematical uncertainty of the fitting routine.
The missing points are due to the low intensity and a not converging fit for some wavelength.
The data points extracted from the (-111) and (022) reflections are shifted by 10 and 20 nm, respectively, for reasons of clarity.
As seen from the figure no change in lattice parameter with depth as well as no tetragonal distortion is found, even though the correlation length and epitaxy of the structures formed at the two interfaces differs largely.
However, for the very low penetration depth the position of the Bragg reflections is difficult to measure since the scattered intensity is extremely low and the resolution, as discussed in the next section, is determined by the angular divergence of the incident beam.
Regarding the uncertainty in the determination of the lattice parameter and the fact that the Q value at the position of the Bragg reflections is more than three times the critical momentum transfer we have not corrected the data for refraction effects.

\subsection{Resolution}

From figure \ref{scattgeo}, right panel, it is seen that the penetration depth into the liquid changes by three orders of magnitude within 10 \% above and below $Q_c$.
This implies that in order to reach a good depth resolution an excellent $\frac{\Delta Q}{Q_c}$ is needed.
$\frac{\Delta Q}{Q}$ is calculated by the following equation taking into account the wavelength distribution $\Delta \lambda$ and a angular divergence $\Delta \theta$:
\begin{equation}
\frac{\Delta Q}{Q} = \sqrt{\left( \frac{\Delta \lambda}{\lambda} \right)^2 + \left( \Delta \theta \cot \theta \right)^2}
\end{equation}
For small angles $\theta$ this can be simplified:
\begin{equation}
\frac{\Delta Q}{Q} = \sqrt{\left( \frac{\Delta \lambda}{\lambda} \right)^2 + \left( \frac{\Delta \theta}{\theta} \right)^2}
\end{equation}
For a time of flight instrument the wavelength resolution is given by the length of the instrument and the time resolution of the detector.
If the acquisition is run in event mode the wavelength resolution can be chosen after the experiment and be optimized to the signal.
However, the lower limit in the wavelength resolution is given by the length of the neutron pulse $\Delta t$ and the flight time from the source to the detector and can be calculated from:
\begin{equation}
\Delta \lambda = \frac{h \Delta t}{m_n L}
\end{equation}
with $h$, $m_n$ and $L$ being Plancks number, the mass of the neutron and the distance from the source, which is the cold source at TS2 at ISIS for the present case, to the detector, respectively.
Plugging in the numbers for $L = 25$ m and $\Delta t$, which is between 0.3 and 0.6 ms, results in an intrinsic wavelength resolution of about 1 \% or  5 pm at the critical wavelength which is close to 0.53 nm for the angular settings chosen here.
On the other hand the resolution in angular divergence is given by the settings of the collimation slits and was 0.04$^\circ$ in the present study.
The incident beam angle was 0.3$^\circ$ resulting in a relative uncertainty of 13.3 \%.
However, even though the angular resolution was relaxed the intensity for wavelengths larger than the critical wavelength were not sufficient to extract a signal from the very near surface layer.
To allow such studies much longer counting times or stronger pulsed neutron sources are needed.\\

\section{Conclusion}

In this paper we present a TOF-GISANS experiment to resolve the near surface structure of micellar systems.
By use of TOF neutron reflectometry and small angle scattering depth sensitive information is extracted in one single measurement.
It turns out that with increasing distance from the interface the crystallographic structure approaches the bulk structure.
Our experiments open new possibilities for the investigation of near-surface structures over several orders in length scale not only of micellar systems but also of colloids.

\section{Acknowledgement}

The authors thank Katharina Theis-Br\"ohl for sending Georg Liesche for a project based at the Institute Laue-Langevin and Peter Kuhns for help with the measurements at ISIS.
In addition we acknowledge financial support from the Swedish research council VR under contract number A0505501 and the European Union NMI-3 initiative for financial assistance for travel as well as the STFC for the beam time, RB1120261.

\bibliography{RheoRefl}

\end{document}